\documentclass[]{scrartcl}
\usepackage{graphics}
\usepackage[english]{babel}
\usepackage{amsmath} 
\usepackage{amsfonts}
\usepackage{latexsym}
\usepackage{cuted}
\usepackage{graphicx}
\usepackage{graphics}
\usepackage{float}
\usepackage{breqn}
\usepackage{amssymb}
\usepackage{authblk}
\usepackage{multirow}
\usepackage{makeidx}
\usepackage{dblfloatfix}
\usepackage[section]{placeins}
\usepackage{hyperref}
\usepackage{url}
\usepackage{flushend}
\usepackage{microtype}
\PassOptionsToPackage{hyphens}{url}

\begin{document}
\title{A new laser-ranged satellite
for General Relativity and Space Geodesy
}
\subtitle{II. Monte Carlo Simulations and covariance analyses of the LARES 2 Experiment.
}

\author[1,2]{Ignazio Ciufolini\thanks{ignazio.ciufolini@unisalento.it}}
\author[3]{Erricos C. Pavlis}
\author[4]{Giampiero Sindoni}
\author[5]{John Ries}
\author[4]{Antonio Paolozzi}
\author[6]{Rolf Koenig}
\author[7]{Richard Matzner}

\affil[1]{\footnotesize Dip. Ingegneria dell'Innovazione, Universit\`a del Salento, Lecce, Italy}
\affil[2]{Museo della fisica e Centro studi e ricerche Enrico Fermi, Rome, Italy}
\affil[3]{Joint Center for Earth Systems Technology (JCET), University of Maryland, Baltimore County, USA}
\affil[4]{Scuola di Ingegneria Aerospaziale, Sapienza Universit\`a di Roma, Italy}
\affil[5]{Center for Space Research, University of Texas at Austin, Austin, USA}
\affil[6]{Helmholtz Centre Potsdam, GFZ German Research Centre for Geosciences, Potsdam, Germany}
\affil[7]{Theory Center, University of Texas at Austin, Austin, USA}

\renewcommand\Authands{ and }

\date{}
\maketitle
\abstract{
In the previous paper we have introduced the LARES 2 space experiment. The LARES 2 laser-ranged satellite is planned for a launch in 2019 with the new VEGA C launch vehicle of the Italian Space Agency (ASI), ESA and ELV. The main objectives of the LARES 2 experiment are gravitational and fundamental physics including accurate measurements in General Relativity, and accurate determinations in space geodesy and geodynamics. In particular LARES 2 is aimed to achieve a very accurate test of frame-dragging, an intriguing phenomenon predicted by General Relativity. Here we report the results of Monte Carlo simulations and covariance analyses fully confirming an error budget of a few parts in one thousand in the measurement of frame-dragging with LARES 2 as calculated in our previous paper.
}
\section{Introduction}
\label{intro}
The LARES 2 satellite is a laser ranged satellite aimed at achieving a test of frame-dragging\cite{bib1ter,bib1quad} \footnote{Also called the Lense-Thirring effect.}, an intriguing phenomenon predicted by General Relativity, with an accuracy of a few parts per thousand. It is planned for a launch in 2019 with the new VEGA C launch vehicle of the Italian Space Agency (ASI), ESA and ELV. The satellite-ranging data of LARES 2 will be coupled to those of NASA's laser-ranged satellite LAGEOS which has been orbiting the Earth since 1976. The tracking of the two satellites will be provided by the International Laser Ranging Service (ILRS). The idea of the experiment, as described in \cite{bib1bis} is to have two laser-ranged satellites with the same semimajor axis but supplementary inclination in order to eliminate the uncertainties due to the non-sphericity of the Earth's gravitational field and thus to very accurately measure frame-dragging. ({\itshape Supplementary inclinations:} $i_{LARES 2} + i_{LAGEOS} = 180^\circ$.)

In \cite{bib1bis} we showed that LARES 2 will be able to achieve a test of frame-dragging with accuracy of a few parts in a thousand. Here, with a number of Monte Carlo simulations and with a covariance analysis, we show that LARES 2 can indeed achieve a test of frame-dragging with that accuracy. We designed the Monte Carlo simulations and covariance analyses to reproduce as closely as possible the real experiment to measure frame-dragging using LARES 2, LAGEOS, and the GRACE Earth gravitational field determinations.

\section{Design of the Monte Carlo simulations}

The Monte Carlo simulations are performed as follows. The {\itshape first} step is to identify a set of physical parameters whose uncertainties have a critical impact on the accuracy of the measurement of the frame-dragging effect using LARES 2 and LAGEOS. Then, we consider the values of these critical parameters, determined either by the GRACE space mission (in the case of the Earth gravitational field parameters) or by previous extensive orbital analyses (in the case of the radiation pressure parameters of the satellites). Together with the values of these parameters, we consider their realistic 1-$\sigma$ uncertainty estimated by also taking into account the systematic errors.

The Earth gravitational field model that is considered is GOCO05s \cite{bib2bis}, a global gravitational field model to degree and order 280. It is estimated from data of the satellite gravity missions GOCE, GRACE and CHAMP. The values of the critical parameters and of their 1-$\sigma$ uncertainties are given in Table 1. An exceptional entry in Table 1 is $C_{2,0}$, the Earth's quadrupole coefficient. It is obtained from Satellite Laser Ranging \cite{bib2ter} and has a significant time dependence. Section 5 below discusses our assessment of its value and uncertainty. (We take the $1-\sigma$ uncertainty as $0.5 \times 10^{-11}$.)
In \cite{bib1bis} and in a number of papers describing the details and the error analysis of the LAGEOS 3 experiment \cite{bib3bis,bib4bis,bib5bis,bib6bis,bib8bis} see also \cite{bib9bis,bib10bis,bib11bis,bib12bis,bib13bis}, we described the main error sources in the measurement of frame-dragging using two laser-ranged satellites with supplementary inclinations. In succeeding papers \cite{bib14bis,bib14ter} we treat the errors due to the de Sitter effect (geodetic precession) and thermal drag.

\begin{table}
  \centering
  \begin{tabular}{|l|l|l|}
     \hline
   Parameter&	Nominal Value&	1-sigma uncertainty\\
GM$_{\bigoplus}$	 &0.3986004415 $\cdot$ 10$^{15}$  m$^{3}$/s$^{2} $ &	8$\cdot$10$^{5}$ m$^{3}$/s$^{2}$ \\
C$_{2,0}$&	-0.48416521$\cdot$10$^{-3}$	&0.5 $\cdot$10$^{-11}$\\
C$_{4,0}$&	 0. 539998$\cdot$10$^{-6}$	&0.0614 $\cdot$10$^{-11}$\\
C$_{6,0}$&	-0. 149975$\cdot$10$^{-6}$	&0.36515$\cdot$10$^{-12}$\\
C$_{8,0}$&	 0. 49477$\cdot$10$^{-7}$	&0.26795$\cdot$10$^{-12}$\\
C$_{10,0}$&	 0. 53342$\cdot$10$^{-7}$	&0.2189$\cdot$10$^{-12}$\\
\.{C}$_{2,0}$&	 1.207$\cdot$10$^{-11}$	&0.00895$\cdot$10$^{-11}$\\
\.{C}$_{4,0}$&	 0.47$\cdot$10$^{-11}$	&0.165$\cdot$10$^{-12}$\\
C$_{3,0}$&	 9.57173$\cdot$10$^{-7}$	&0.6531$\cdot$10$^{-11}$\\
C$_{5,0}$&	 6.8646$\cdot$10$^{-8}$	&1.61115$\cdot$10$^{-12}$\\
C$_{r}$ LAGEOS  &	 1.13&	0.3$\cdot$10$^{-2}$\\
C$_{r}$ LARES 2&	 1.10&	0.3$\cdot$10$^{-2}$\\

     \hline
   \end{tabular}
  \caption{The parameters considered in the Monte Carlo simulations with their 1-sigma uncertainties}\label{table1}
\end{table}

The value of GM$_{\bigoplus}$, the  gravitational constant times the Earth mass, used herein is consistent with GOCO05S, i.e., is set by definition in the GOCO05s gravity field model development. Its value and standard deviation are taken from the International Earth Rotation and Reference Systems Service (IERS) Conventions 2010, where the current knowledge of this parameter is documented. The standard deviations of the values of the first few even zonal harmonics of the gravity field model, i.e. C$_{2,0}$, C$_{4,0}$, C$_{6,0}$, C$_{8,0}$, C$_{10,0}$, of the secular rate of change of the two largest even zonal harmonics, \.{C}$_{2,0}$ and \.{C}$_{4,0}$, and of the odd zonal harmonics C$_{3,0}$ and C$_{5,0}$, are calibrated values i.e., including the estimated systematic errors. The solar radiation coefficient, C$_{r}$, of the LAGEOS satellite is taken from long-term use in the geodetic community, while that for LARES 2 has been extrapolated form the values of the C$_{r}$ of LAGEOS and LARES.

The {\itshape second} step is to randomly generate 15 samples of values for each parameter of Table 1, with population distributed as a normal (Gaussian) distribution around the mean value of each parameter that is equal to its value reported in Table 1, and with standard deviation equal to the calibrated sigma of that parameter, also reported in Table 1. Then 10 sets of initial conditions for the LARES 2 orbital elements were randomly generated by considering the sigmas taken from the VEGA manual and confirmed by ELV, to take into account the estimated orbital injection uncertainties of VEGA C.

The {\itshape third} step is to generate the orbits of LARES 2 and LAGEOS. This is done by using the orbital propagator and estimator GEODYN, using each time, as input, the values of one of the 150 sets of the sixteen parameters that were generated at the second step. The frame-dragging effect is always kept equal to its calculated General Relativity value. In the end, 150 different cases for the evaluation of the frame-dragging effect using LARES 2 and LAGEOS were obtained.
 These 150 simulations represent 150 approximations of the real orbit of LAGEOS and of the proposed orbit of LARES 2, generated by physical perturbations that are partially unknown because of the uncertainties in the parameters of Table 1. In a second orbital propagation, we generate the orbit of each satellite, starting with the same initial conditions of the previous case but using the nominal value of each of the considered parameters and zero frame-dragging effect. This second set of simulations of the reference orbit of each satellite represents the set of orbits of each satellite as modeled in the orbital data analyses using the orbital estimators of the LARES team, i.e., GEODYN, EPOS-OC and UTOPIA.

Finally, for each 15-day arc, we take the difference of the nodes between these two sets of orbits, i.e., between each one of the 150 cases described above and the nominal case obtained using GEODYN, and save that difference for each 15-day arc and for each satellite. In this way, for over a period of 1035 days, almost the period of the node (1050 days), we obtain 150 sets of 69 simulated residuals for each satellite, representing the noisy residuals of each satellite that will be obtained in the real data analysis. These 150 sets of simulated nodal residuals for each satellite provide a reasonably large number of cases according to the Central Limit theorem.

In summary, for each of the 150 simulations we generate 69 simulated residuals for each of the satellites, LARES 2 and LAGEOS. Each of these simulations, with the corresponding residuals, is obtained with one of the sets of physical parameters generated according to the mean and sigma reported in Table 1.

Then, the node residuals of LARES 2 and LAGEOS are combined in order to eliminate the effect on their nodes of the even zonal harmonics and of their variations   \cite{bib3bis,bib4bis,bib5bis,bib6bis,bib8bis,bib9bis,bib10bis,bib11bis}. We combine the residuals of each randomly generated orbit of LARES 2 with the residuals of an orbit of LAGEOS generated using the same random set of geophysical parameters GM$_{\bigoplus}$, C$_{2,0}$, C$_{4,0}$, C$_{6,0}$, C$_{8,0}$, C$_{10,0}$, \.{C}$_{2,0}$, \.{C}$_{4,0}$, C$_{3,0}$ and C$_{5,0}$. For each of the 150 simulations the 69 combined residuals are integrated to get the cumulative residual shift of the combination of LARES 2 and LAGEOS. Finally, the combination of the integrated residuals of the two satellites is fitted with a straight line using the least squares method, to obtain, for each of the 150 simulations, the simulated, measured value of the frame-dragging effect.

Figures 1 and 2 show the cumulative residual shift of the node of LAGEOS and LARES 2 and their combination respectively, for each of the 150 Monte Carlo simulations.

\section{Results of the Monte Carlo simulations}

Figure 1  shows the simulated nodal drifts corresponding to each of the 150 simulations for LAGEOS and LARES 2. Figures 2 and 3 refer to the LAGEOS and LARES 2 combination. Each nodal drift shown in Figure 2 was obtained by combining the simulated cumulative (integrated) node residuals of the two satellites for each of the 150 simulations and by fitting the raw residuals with a straight line. The result of the 150 simulations for the LARES 2 and LAGEOS combination is that the mean value of the measured frame-dragging effect is equal to 100.016\% of the frame-dragging effect modeled in GEODYN (i.e. there is a deviation of 1.6 $\cdot$ 10$^{-4}$ with respect to frame-dragging), with a standard deviation equal to 1.3 $\cdot$  10$^{-3}$ of the frame-dragging effect modeled in GEODYN, i.e. the frame-dragging (or Lense-Thirring) parameter $\mu$ (set up equal to 1 corresponding to General Relativity), is:
\begin{equation}\label{eq1}
 \mu = 1.00016 \pm 0.0013
\end{equation}

\noindent The 1.3 $\cdot$  10$^{-3}$  uncertainty represents the systematic errors in the measurement of frame-dragging with the LARES 2 experiment. This result fully agrees with the error analysis of the LARES 2 experiment reported in our previous paper \cite{bib1bis}.

\begin{figure}
\centering
 \includegraphics[width=0.940\textwidth]{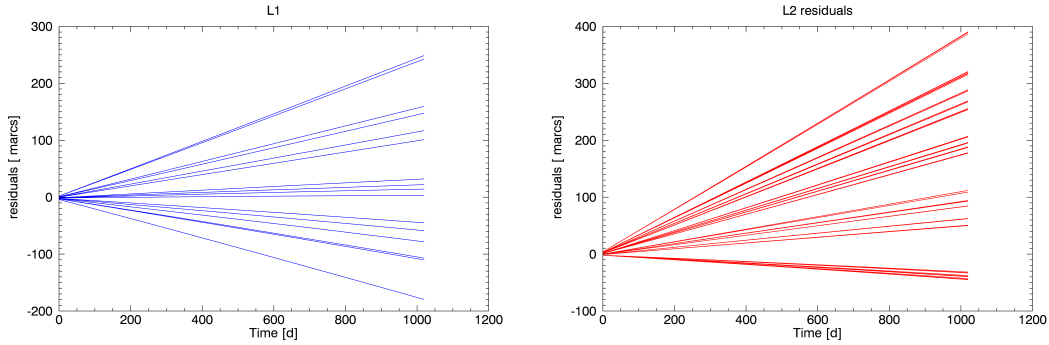}
\caption{Left: Simulated cumulative (integrated) residuals of the nodal longitude for each of the 150 performed Monte Carlo simulations for the LAGEOS satellite. Right: Simulated cumulative residuals for LARES2.}
\label{fig:1}       
\end{figure}

\begin{figure}
\centering
 \includegraphics[width=0.640\textwidth]{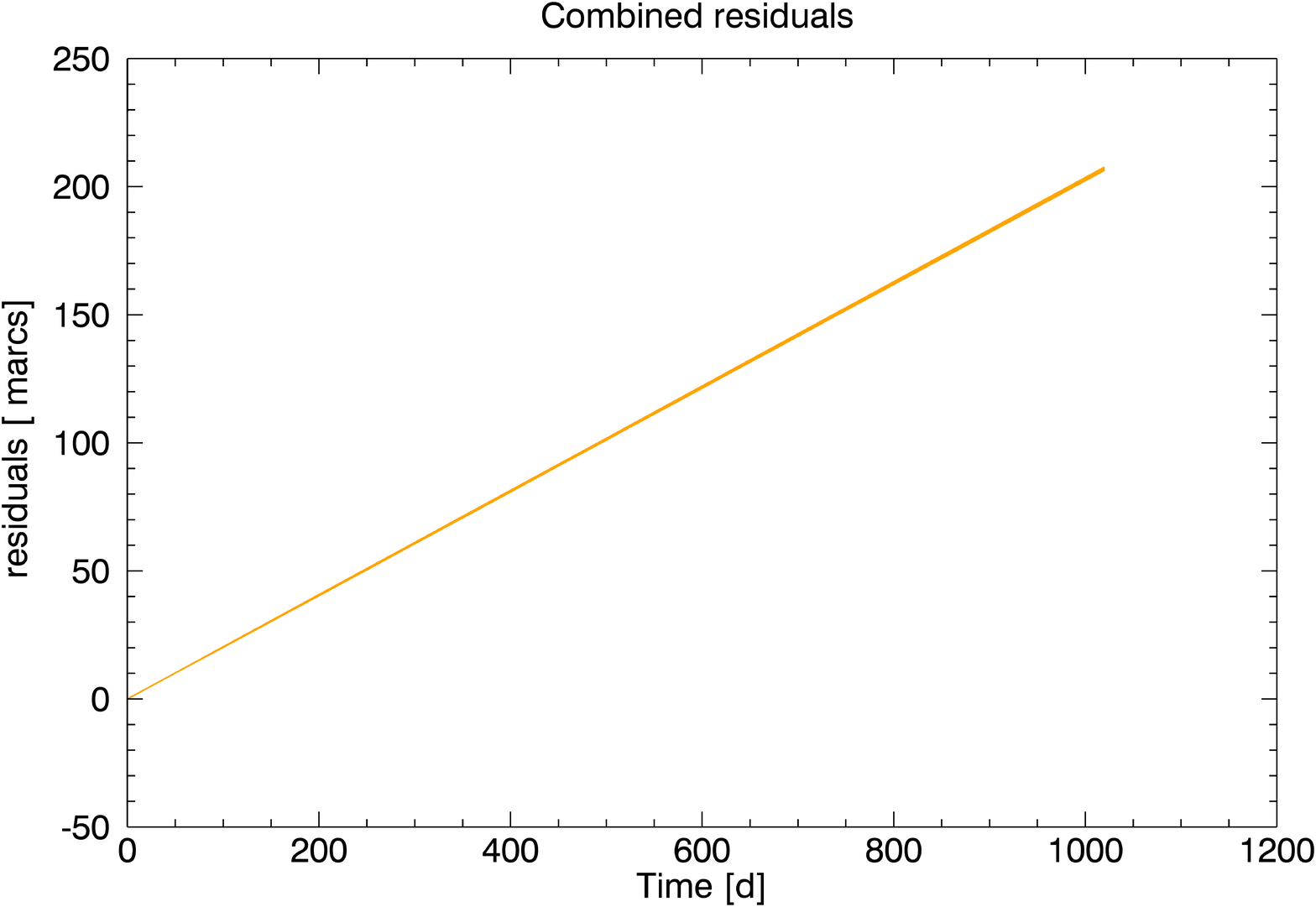}
\caption{Simulated cumulative trend of residuals of the combination of the nodal longitudes of     LAGEOS and LARES 2, for each of the 150 performed Monte Carlo simulations. To the accuracy of this graph the results are {\itshape all} identical, and identical to the General Relativistic Lense-Thirring  prediction. See Eq (1).}
\label{fig:2}       
\end{figure}

\begin{figure}
\centering
 \includegraphics[width=0.640\textwidth]{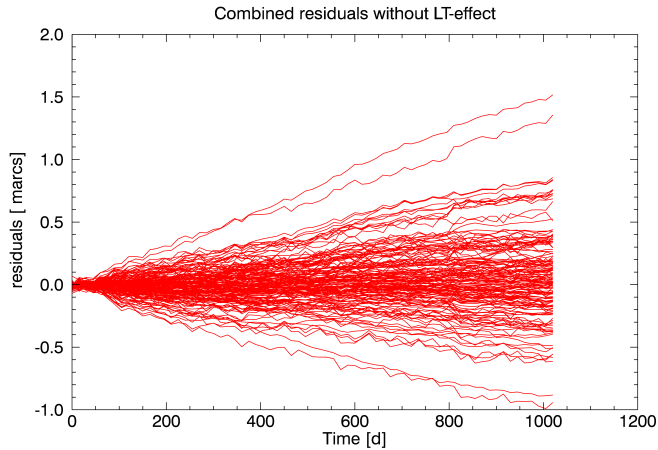}
\caption{Simulated cumulative residuals of the combination of the nodal longitudes of LARES 2 and LAGEOS, for each of the 150 performed Monte Carlo simulations, after removing the trend due the Lense-Thirring effect predicted by General Relativity. The range of the y-axis is one hundred times smaller than that of Fig. 2. The average slope of each line results from orbital injection errors; the slopes are very small compared to the frame-dragging effect.}
\label{fig:3}       
\end{figure}

The combination of Figure \ref{fig:3}   eliminates the uncertainties in all the Earth's even zonal harmonics, $C_{2n,0}$.\cite{bib3bis,bib4bis,bib5bis,bib6bis,bib8bis,bib9bis,bib10bis,bib11bis} Figure \ref{fig:3} clearly displays the reduction of the spread between the 150 simulations, i.e., the reduction of the standard deviation of the slopes of the 150 simulations, when all the $C_{2n,0}$ uncertainties are removed from the residual nodal drifts of the satellites by using the combination of their nodal residuals. The use of LARES 2 with a supplementary orbit to that of LAGEOS dramatically reduces the standard deviation of the slopes of the nodal residuals of the 150 simulations, that is, it reduces the uncertainty in the simulated measurement of frame-dragging using LARES 2 and LAGEOS.

\section{Covariance Analysis of the LARES 2 Experiment}

We have also used standard covariance analysis to propagate the errors of ``considered'' parameters (i.e. held fixed, not estimated on the basis of SLR data) on the main parameter of interest for this experiment, the frame-dragging, or Lense-Thirring parameter $\mu$ (set up equal to 1 corresponding to General Relativity).

Our estimation of $\mu$ is obtained from the combination of the nodal residuals of the orbits of LAGEOS and LARES-2. Therefore, the errors affecting a number of ``considered" parameters need to be propagated on these elements and then the combination must be performed in the way we have outlined in our previous paper \cite{bib1bis}. The list of ``considered" parameters and the adopted errors is the same as that used for the Monte Carlo approach, so that the results from the two approaches are totally consistent and comparable.

The theoretical basis of the covariance analysis can be found in any standard textbook of statistics (e.g.\cite{bib15bis}) . The errors in the parameters x = (x$_{1}$, x$_{2}$, ..., x$_{n}$) of a function

\begin{equation}\label{f2}
  f=f(x_{1}, x_{2}, ..., x_{n})=Ax
\end{equation}

\noindent described by a covariance $\Sigma_{x}$ are propagated onto the function $f$ using the combination coefficients $A_{i}$ that form the ``design'' matrix $A$. If the function $f$ is non-linear, as is the case here, then we can still use this approach, but we need to modify it by inserting a linearization of $f$ using a Taylor series expansion, e.g.:
\begin{equation}\label{f3}
  f\approx f^{0} + Jx
\end{equation}

\noindent where $ f^{0}$ is the function evaluated at $ x_{0}$  and $J$ is the matrix of partial derivatives of $f$ with respect to $x$, known also as the Jacobian matrix. Each component of $J$ is:
\begin{equation}\label{f4}
  J_{ij}= \partial f_{i}/ \partial x_{j}
\end{equation}

\noindent and the propagation can be performed on the part of $f$ that depends on $x$ since $f_{0}$ is a constant and not affected by errors, using the quadrature formula:
\begin{equation}\label{f5}
  \Sigma_{f}=J \Sigma_{x} J^{T}
\end{equation}

These are the exact formulae that are used in our estimation of the parameters from our GEODYN estimator, where $f$ is now the state vectors of the two missions of interest here, LAGEOS and LARES-2.

This same process can be slightly modified to accommodate the ``considered'' parameters and their errors in order to ``map'' their effect on the main unknowns of our problem, the two state vectors and eventually the final parameter $\mu$. This is accomplished by introducing all of the ``considered'' parameters as formal parameters in each arc, albeit with very small a priori sigma so that they are actually not adjusted at all, but their partial derivatives (components of the Jacobian) are formed and included in the normal equations (NEQs) used for the solution of the problem for all parameters ``X''. The vector X consists of two groups of parameters:
\begin{equation}\label{f6}
 X = [ y : x ]
\end{equation}

\noindent the vector y comprises the standard parameters that we will adjust (solve for) based on the SLR data. The second group of parameters x are the ``considered'' parameters, the ones for which we only want to propagate their errors onto the solution for the y subset of parameters.

We accomplish this with a second piece of NASA Goddard software, the linear equation solver ``SOLVE'' \cite{bib16bis} . The simulated data are used in GEODYN to generate a set of NEQs for each arc using the nominal values of the parameters ``x''. We then use these NEQs in SOLVE, where we can discriminate between parameters whether in groups or individually, defining which are to be solved for (as in a standard solution for the orbital arcs) and which we only need to have their errors propagated. For the second set their values are ``shifted'' using the partial derivatives in the Jacobian, to the new values that contain the errors of our choice. The NEQs are then solved for the ``y'' subset of the parameters based on the modified values of the ``x'' ``considered'' parameters. The adjusted estimates for ``y'' and their covariance contain the effect of the errors applied on the ``x'', ``considered'' parameters.

The final step is the propagation of this effect onto $\mu$ and this is accomplished by re-computing the individual arcs’ orbits based on the modified ``considered'' parameters and the corresponding adjusted estimates of the state vectors. Similar to the Monte Carlo approach, we need to repeat this for each arc several times using (randomly) different errors for the ``considered'' parameters to avoid basing our decision on a single realization of the errors. By the central limit theorem, the average of a sufficiently large number of replications converges to the true estimate.

In our experiment, we performed 20 trials of the variation of the ``considered'' parameters. That resulted in twenty distinct estimates of the LAGEOS and LARES-2 state vectors, leading to 20 corresponding orbital series (see figure 4 with the cumulative node residuals of LAGEOS and LARES 2), that were subsequently combined to produce 20 estimates of $\mu$ (see figure 5). With the new SLR systems operating at kHz rates and providing tremendous numbers of raw ranges the normal points (NP) that we use in our adjustments are totally free of measurement noise due to the averaging process to form the NP. We are thus not limited by such errors but rather by systematic errors in the other parameters that we cannot improve by adjusting them simultaneously from the same SLR data, i.e. those that we choose here to be the ``considered'' parameters.

The mean and scatter of these 20 estimates of $\mu$ are due to the variation of the ``considered'' parameters from the errors we applied and they are a considerably more reliable estimate of the error incurred by $\mu$ than something that we could compute using some theoretical formulation or the formal statistics of an adjustment of SLR data. The result of the covariance analysis is:
\begin{equation}\label{f7}
\mu=1.0007\pm 0.0019
\end{equation}
	\begin{figure}
\centering
 \includegraphics[width=0.640\textwidth]{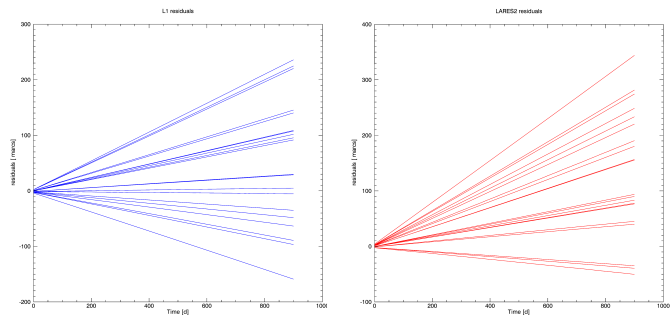}
\caption{Result from 20 covariance analysis realizations: Cumulative node residuals of LAGEOS (left) and LARES2 (right). }
\label{fig:4}       
\end{figure}				
\begin{figure}
\centering
 \includegraphics[width=0.640\textwidth]{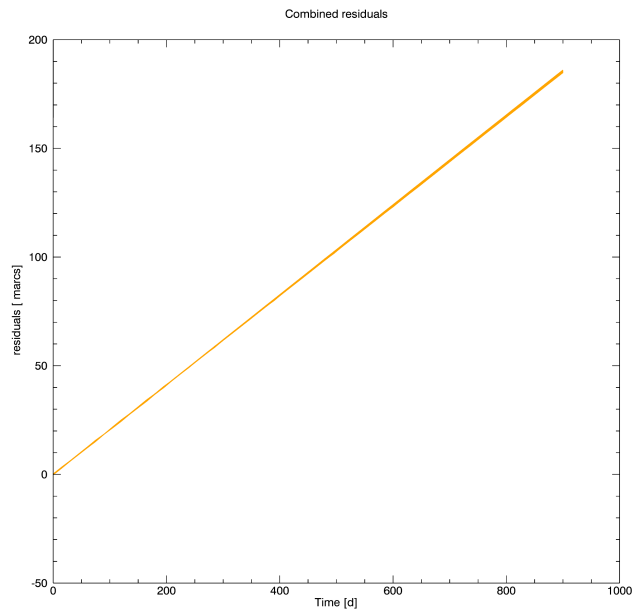}
\caption{Result from 20 covariance analysis realizations: Combined cumulative node residuals of LAGEOS and LARES 2. }
\label{fig:5}       
\end{figure}
In conclusion the covariance analysis has confirmed the results of the Monte Carlo simulations presented in the previous section 3 and the error budget estimate provided in \cite{bib1bis}. However, the Monte Carlo analysis was carried out over 150 cases whereas the covariance analysis was carried out over only 20 cases due to computational time limitations.

\section{The Earth quadrupole moment and its uncertainty in the LARES 2 experiment}

The value for $C_{2,0}$ in GGM05S is actually from Satellite Laser Ranging (SLR), based on the paper \cite{bib17bis}. In this section we discuss the value and the uncertainty of the Earth quadrupole moment, measured by the normalized $C_{2,0}$ coefficient, in terms of the equivalent non-normalized Earth quadrupole coefficient $J_{2}$ = - $\sqrt{5}$  $C_{2,0}$ .
The value in GGM05S is the quadratic (see Figure 6) described in \cite{bib17bis}, evaluated at epoch 2008 (the approximate midpoint of the GRACE data used to determine GGM05S). The error estimate in GGM05S was deliberately conservative, because there is no reliable way to calibrate the `mean' value of $C_{2,0}$ in the presence of such a large long-term variation.

\begin{figure}
\centering
 \includegraphics[width=0.640\textwidth]{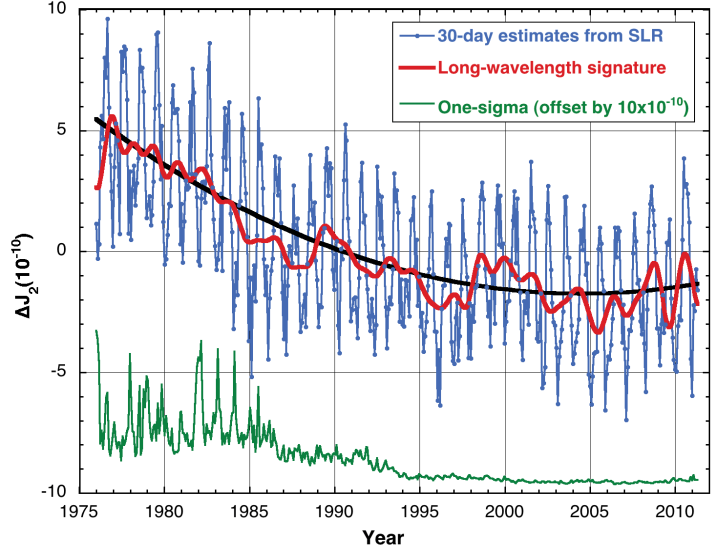}
\caption{Adapted from \cite{bib17bis}. 30 day estimates of J$_{2}$ from SLR (blue line) and its long-wavelength signature represented by the decadal
 spectral band of the wavelet filtering (red line). The uncertainty estimates (green line) are offset by 10 $\cdot$ 10$^{−10}$ for clarity. Superposed is a
quadratic fit (black line) to the 30 day estimates illustrating the quadratic nature of the long-term trend. }
\label{fig:6}       
\end{figure}
In \cite{bib17bis}, Cheng, Tapley and Ries assigned an uncertainty to the value of $J_{2}$ at epoch 2000 of 1 $\cdot$ 10$^{-11}$, based on the results of a quadratic fit to the $J_{2}$ time series.

This is in line with the following argument about the ``uncertainty" in $J_{2}$ from SLR regarding the inclination accuracy. As \cite{bib17bis} shows (c.f. Figure 6), the monthly $J_{2}$ error estimates are about 15 $\cdot$  10$^{-11}$ (ignoring the early days). Similarly, see the current Release-05 version of the UT/CSR degree-2 time series at \url{ftp://ftp.csr.utexas.edu/pub/slr/degree_2/C20_RL05.txt} where we see that the $C_{2,0}$ error estimates range from 3 to 5 $\cdot$ 10$^{-11}$, which is about 6.7 to 11 $\cdot$  10$^{-11}$ for $J_{2}$. The degree-2 time series may be a little bit more accurately calibrated, but we can say with some confidence that the uncertainty for $J_{2}$ is around 10$^{-10}$. Since this is the uncertainty for each monthly value, the accumulated error in the experiment would benefit from root($N$) where $N$ represents the number of months in the solution.

So if we use a model for $J_{2}$ based on the monthly estimates, the ``effective uncertainty", i.e. the overall impact of $J_{2}$ on the Lense-Thirring experiment due to inclination error, would be reduced by roughly root-N, where N are the number of months in the experiment. This leads to something approaching 10$^{-11}$ when we have several years of data available for the experiment. As long as we do not treat $J_{2}$ as constant but rather recognize and model the long-term variation, the error is smaller than simply treating $J_{2}$ as a constant.

As Figure 6 shows, we cannot characterize $J_{2}$ by a mean value, but we will have monthly estimates of $J_{2}$ indefinitely into the future with comparable accuracy. In fact, with the addition of LARES and LARES-2 to the mix, $J_{2}$ may be estimated more accurately. In any case, with $\sim$ 100 months of data, we could conclude that the overall effect of $J_{2}$ on the experiment through the orbit injection error would be reduced to the equivalent of 10$^{-11}$, that is about 0.4 $\cdot$  10$^{-11}$ for $C_{2,0}$. We use 0.5 $\cdot$  10$^{-11}$ for $C_{2,0}$ in our analyses.

\section{Summary and Conclusions}

The combination of the observables provided by the nodes of the two satellites LARES 2 and  LAGEOS, with supplementary inclinations, will allow a test of the phenomenon of frame-dragging predicted by General Relativity with an  uncertainty of a few parts per thousand, by eliminating the uncertainties in all the even zonal harmonics of the Earth potential $C_{2n,0}$.  The analyses outlined in \cite{bib1bis,bib14bis,bib14ter} based on a number of previous detailed and extensive error analyses \cite{bib3bis,bib4bis,bib5bis,bib6bis,bib8bis,bib9bis,bib10bis,bib11bis,bib12bis,bib13bis}, have confirmed that the LARES 2 experiment can achieve a measurement of frame-dragging with such uncertainty. Nevertheless, to further test the previous extensive error analyses, we have designed and performed 150 Monte Carlo simulations of the LARES 2 and LAGEOS experiment. In our Monte Carlo analysis we have simulated the orbits of the LARES 2 and LAGEOS satellites by randomly generating the values of the GM$_{\bigoplus}$ (mass) of Earth, of its five largest even zonal harmonics, $C_{2,0}$, $C_{4,0}$, $C_{6,0}$, $C_{8,0}$,$C_{10,0}$, of the secular rate of change of the two largest even zonal harmonics, $\dot{C}_{2,0}$ and $\dot{C}_{4,0}$, and of the odd zonal harmonics $C_{3,0}$ and $C_{5,0}$, and of the solar radiation coefficients of LARES 2 and LAGEOS 2. These parameters are identified as the main source of bias in the measurement of frame-dragging using LARES 2 and LAGEOS. The LARES 2 orbits were also simulated taking into account the injection accuracy of VEGA C. The mean of the frame-dragging effect measured in the 150 simulations was equal to 100.016\% of the frame-dragging effect modeled in GEODYN. The standard deviation of the frame-dragging effect measured in the 150 simulations, representing the systematic errors in the measurement of frame-dragging, was 1.3 $\cdot$ 10$^{-3}$ of the combined frame-dragging effect. The covariance analysis has substantially reproduced the results of the Monte Carlo analysis and the error budget estimate provided in the original LARES 2 draft proposal. However, the Monte Carlo analysis was carried out over 150 cases whereas the covariance analysis was carried out over only 20 cases.

\end{document}